\documentclass[a4paper]{article}

\usepackage{INTERSPEECH2020}
\usepackage{float}

\title{Formant Tracking Using Dilated Convolutional Networks Through Dense Connection with Gating Mechanism}
\name{Wang Dai$^1$, Jinsong Zhang$^1$, Yingming Gao$^2$, Wei Wei$^1$, Dengfeng Ke$^3$, Binghuai Lin$^4$, Yanlu Xie$^1$}
\address{
  $^1$School of Information Science, Beijing Language and Culture University, China\\
  $^2$Institute of Acoustics and Speech Communication, TU Dresden, Germany\\
  $^3$National Laboratory of Pattern Recognition, Institute of Automation, Chinese Academy of Sciences, China\\
  $^4$MIG, Tencent Science and Technology Ltd., Beijing, China}
  
\email{daiwang\_ai@163.com, \{jinsong.zhang,xieyanlu\}@blcu.edu.cn,\\
yingming.gao@mailbox.tu-dresden.de, wwei906@163.com,\\
 dengfeng.ke@nlpr.ia.ac.cn, binghuailin@tencent.com}
\begin{document}

\maketitle
\begin{abstract}
Formant tracking is one of the most fundamental problems in speech processing. Traditionally, formants are estimated using signal processing methods. Recent studies showed that generic convolutional architectures can outperform recurrent networks on temporal tasks such as speech synthesis and machine translation. In this paper, we explored the use of Temporal Convolutional Network (TCN) for formant tracking. In addition to the conventional implementation, we modified the architecture from three aspects. First, we turned off the ``causal'' mode of dilated convolution, making the dilated convolution see the future speech frames. Second, each hidden layer reused the output information from \textit{all} the previous layers through dense connection. Third, we also adopted a gating mechanism to alleviate the problem of gradient disappearance by selectively forgetting unimportant information. The model was validated on the open access formant database VTR. The experiment showed that our proposed model was easy to converge and achieved an overall mean absolute percent error (MAPE) of 8.2\% on speech-labeled frames, compared to three competitive baselines of 9.4\% (LSTM), 9.1\% (Bi-LSTM) and 8.9\% (TCN).
\end{abstract}


\section{Introduction}
Formants are considered to be resonances of the vocal tract during speech production. An accurate estimation of formant frequencies in spontaneous speech is often desired in many phonological experiments of laboratory phonology, sociolinguistics, and bilingualism \cite{clopper2014effects,munson2004effect}. They also play a key role in the perception of speech and are useful in the coding, synthesis and enhancement of speech, as every phoneme has a unique formants distribution, especially on vowels and sonorous consonants.

Classical formant tracking algorithms are based on peak picking from Linear Predictive Coding (LPC) spectral analysis \cite{mccandless1974algorithm,deng1987composite, steiglitz1977simultaneous,atal1978linear}. LPC spectral coefficients yield intra-frame point estimates of candidate frequency parameters via root finding or peak-picking. The inter-frame parameter selection and smoothing can be performed by minimizing various cost functions in a dynamic programming environment \cite{sjolander2000wavesurfer,boersma2018praat}. However, these classical approaches have an obvious shortcoming that the required root-finding or peak-picking procedure cannot be written in closed form \cite{mehta2012kalman}. More elaborate methods used probabilistic and statistical models to obtain confidence intervals around the estimated formant tracks \cite{mehta2012kalman}, such as quantization of Vocal Track Resonances (VTR) space
\cite{deng2006tracking}, Kalman filtering
\cite{mehta2012kalman,deng2004structured,deng2006adaptive}, HMM \cite{kopec1986formant,lee2005formant,toledano2006initialization} and GMM \cite{darch2006map}.
    
The aforementioned ad-hoc signal processing methods \cite{dissen2016formant} usually emerge false peaks and formant merging when affected by high pitch or coarticulation. These problems can be alleviated by visually correcting with the help of linguistic knowledge and spectral analysis. Motivated by this idea, Deng et al. released a handpicked VTR/Formants corpus in 2006 \cite{deng2006database}. It was subsequently adopted by some researchers as benchmark dataset to develop and evaluate new algorithms for formant tracking. For example, Mehta et al. evaluated their proposed Kalman-based autoregressive moving average modeling methods on this database \cite{mehta2012kalman}. Inspired by the great success of deep learning in many application areas, Dissen et al. employed Long Short-Term Memory (LSTM) networks to train a supervised regression model between LPCCs plus Pitch-Synchronous Cepstrum Coefficients (named PSCCs) and hand-corrected formant frequencies for every speech frame \cite{dissen2016formant}.
Later, Dissen et al. \cite{dissen2019formant} investigated the potential of raw spectrograms (55 $\times$ 50 PSCCs) for formant tracking with Convolutional LSTM networks \cite{xingjian2015convolutional} and found that incorporating the PSCCs and LPCCs achieved the better general performance than using them separately.

Recent studies showed that generic convolutional architectures can outperform recurrent networks on tasks such as speech synthesis and machine translation \cite{amodei2016deep,chan2016listen}. In particular, the Temporal Convolutional Network (TCN) for sequence modeling was proposed \cite{bai2018empirical}, which was composed of dilated causal convolutional networks with residual connection. Stacking convolutional layers with different dilation factors can capture the long-range dependence of the sequence. 
Integrating different hidden features through residual connection makes model more robust. In this work, therefore, we explored whether such advantages of TCN are beneficial for formant tracking. In additon to the application of the conventional TCN model,  we modified its architecture from three aspects: 1) we turned off the ``causal'' mode of dilated convolution, making sure the dilated convolution see the future speech frames; 2) all the dilated convolutions are closely connected, thus effectively reusing the shallow features; 3) we adopted a gating mechanism to automatically select forgetting unimportant information during training. In terms of quantitative error analysis, we compared our proposed approach with other five methods for formant tracking on the VTR test set, including WaveSurfer \cite{sjolander2000wavesurfer}, Praat \cite{boersma2018praat}, LSTM model, Bi-LSTM model and TCN based model.
\begin{figure*}[htb]
  \centering
  \includegraphics[width=\linewidth]{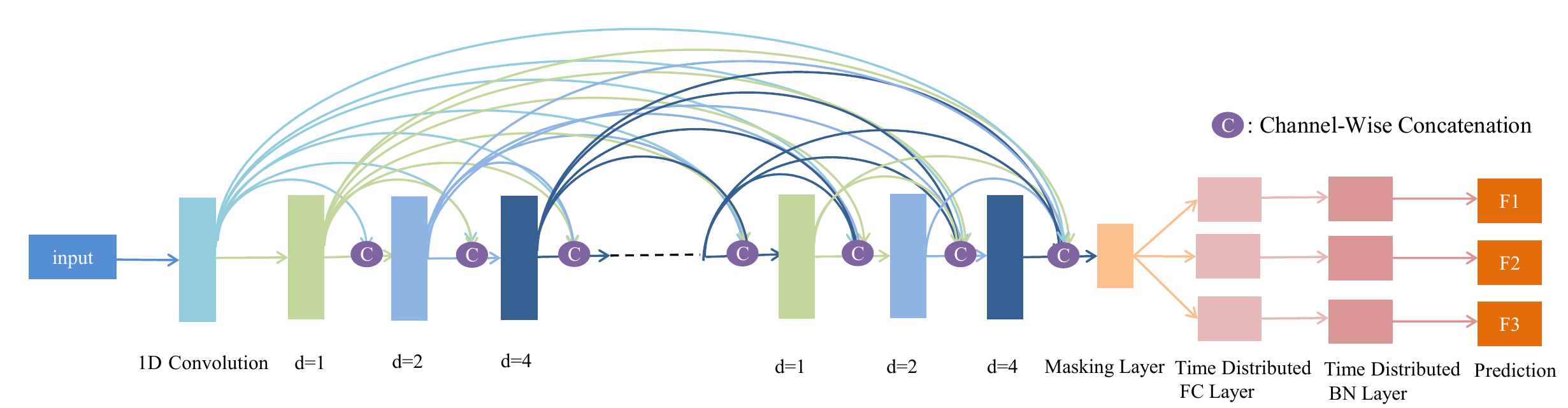}
  \caption{Overview of proposed model framework for formant tracking.}
  \label{fig1:overviewOfFramework1}
\end{figure*}
\begin{figure*}[hbt]
  \centering
  \includegraphics[scale=0.5,height=3cm]{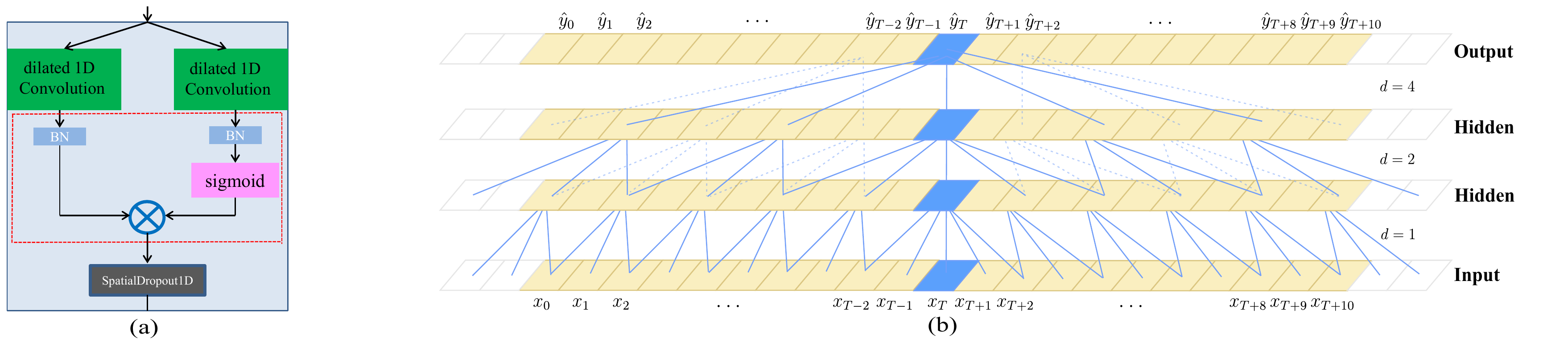}
  \caption{(a) Gated linear unit. (b) An example of dilated 1D convolution with dilation factors $d = 1, 2, 4$ and filter size $k = 3$.}
  \label{fig2:gatingMechanism1}
\end{figure*}
\section{Model Description}
The architecture of proposed model for formant tracking is shown in Figure 1, our framework mainly consists of three components: (1) dilated convolutions, (2) dense connections and (3) gated linear units, all of which are described in subsequent sections.
\subsection{Dilated convolutions}
The generic TCN architecture uses dilated 1D causal convolution where the convolution filter is applied over an area larger than its length by skipping certain input values \cite{bai2018empirical}. Compared with LSTM, the dilated causal convolution needs less nonlinear operations, making the model converge easier. The receptive field can be set to an arbitrary large size by increasing dilation factor. A major disadvantage of dilated causal convolution is the ``causal'' mode when handling the context-dependent sequence tasks. It can only look back at history information because the output at time $T$ is convolved only with elements from earlier to current time step. In speech formant tracking, formants of each frame will be affected by future frames, so we turned off the ``causal'' mode of dilated convolution, making the dilated convolution see the future information. A common practice is to use a dilation factor sequence of form $\{1,2,4,8,...\}$. When $d = 1$, a dilated convolution reduces to a regular convolution. Using larger dilation factor yields an top level output which can capture a wider range of inputs. We provide an illustration in Figure 2(b). In this work, there are 9 1D convolutions stacked with the dilation factors $\{1,2,4,1,2,4,1,2,4\}$ to obtain long context dependence, as shown in Figure 1. Every dilated convolution layer has 64 filters with a size of 3.
\subsection{Dense connections}
The depth of the neural network model is important for learning advanced representations, but it is also accompanied with the challenge of gradient disappearance. Residual training \cite{he2016deep} is considered to be an effective way to address this problem. Using this connection mode, TCN can build a very deep network. Densely connected networks were recently proposed in \cite{huang2017densely}. They can be regarded as a natural evolution version of \cite{he2016deep} where the inputs to a given layer in the network are a concatenation of the outputs from all the previous layers. This way avoids the vanishing gradient problem in depth model. Another advantage is that each layer reused output from all previous layers, such that different level features are fused to improve the robustness of the model. Inspired by the effectiveness of dense connection, we adopted it in our model. A slight difference from \cite{huang2017densely} is that all the dilated convolutions are closely connected to capture more fine-grained features as shown in the densely connected arcs of Figure 1.
\subsection{Gated linear units}
There are several Gating mechanisms that had been explored in modern convolutional architectures for sequential modeling \cite{gehring2017convolutional,dauphin2016language,kalchbrenner2016neural}. Parallel to our work, \cite{dauphin2016language} has shown the form of ($X \times W + b)\ \bigotimes\  \sigma (X \times V + c$) is more effective than others for language modeling. Coupling linear units to the gates, referred to as gated linear units, reduce the vanishing gradient problem. This retains the non-linear capabilities of the layer while allowing the gradient to propagate through the linear unit without scaling. Similarly, in this work, after applying the linearity to the Batch Normalization output of every dilated 1D convolution, we attenuated it with a sigmoid gate (shown in Figure 2(a)). Moreover, we used SpatialDropout1D \cite{tompson2014efficient} at the back of each gated linear unit to sparse the output dimensions (channels) information, thus improving the robustness of the model.

\section{Experiment}
\subsection{Dataset}
VTR corpus \cite{deng2006database} was used in this study to evaluate our model and baselines. It contains 538 SX or SI utterances, selected as a representative subset of TIMIT corpus. Here, SX denotes phonetically compact utterances and SI denotes phonetically diverse utterances. The training set consists of 346, out of which 324 utterances have handpicked VTR. These 346 utterances cover 173 speakers with one SX and one SI utterance from each speaker. The test set consists of 192 utterances covering 24 speakers, and each speaker has 5 SX utterances and 3 SI utterances. Both training and test sets were first annotated by an automatic formant tracking algorithm \cite{deng2004structured}, and subsequently hand-corrected for every 10 ms frame by a group of phonologists based on visual inspection of the first three formants in the spectrogram. We further set aside 24 utterances of 12 speakers (fecd0, mgrl0, falk0, mjrh1, fpaf0, mtrt0, fcdr1, mwsh0, fbch0, msjk0, fjrp1, mdlc1) from the training set as the validation set.
\subsection{Data preprocessing}
Following the study \cite{dissen2019formant}, we used the same acoustic features (LPCCs + PSCCs). A new frame of 30 ms consisted of three original frames of 10 ms in VTR, formant values of which were averaged. In our experiments, we used MATLAB software to extract $8\sim17$ order 30-dim LPCCs (total 300-dim). We first removed the DC component and  applied a pre-emphasis filter ($H(z) = 1 - 0.97z^-{1}$) to the input speech signal. Then the input signal was divided into frames, and the acoustic features were extracted from each frame. The frame shift was 10 ms, and frames were overlapping with Hamming windows of 30 ms. The 50-dim PSCCs was directly extracted from Dissen's open source code \cite{dissen2019formant}.

\subsection{Loss function}
Dissen et al. \cite{dissen2019formant} used a fully connected layer with 3 neurons as the output layer to predict F1, F2 and F3 of each speech frame. The high level features, i.e., the output of the last hidden layer, were shared by each formant but not specific. In fact, there is an inner relationship between formants, each of which has a specific frequency band. Inspired by the success of multitask output \cite{ruder2017overview}, we adopted a similar hard parameter sharing structure. In our framework, there were three parallel branches of fully-connected layers with 256 neurons from dilated convolutional networks. Finally, each of them was linearly transformed to predict the formant. The formant prediction was consider to be independent but mutually restricted to each other in this way. The error between output and reference formant frequency was optimized by the following objective function,

\begin{equation}
  \mathcal{L} = \alpha \times \mathcal{L}_{F1} + \beta \times \mathcal{L}_{F2} + \gamma \times \mathcal{L}_{F3}
  \label{eq:lossFunction}
\end{equation}
where $\mathcal{L}$ is the sum loss of first three formants prediction. $\mathcal{L}_{F1}$, $\mathcal{L}_{F2}$, and $\mathcal{L}_{F3}$ are the losses for F1, F2 and F3, respectively. $\alpha, \beta$, and $\gamma$ represent the weights for the three losses, and they are set to the same value of $1/3$ as each formant prediction deemed to be equally important. To make a fair comparative study, this loss function is applied to all baseline models.

\subsection{Training configuration}
The following experiment settings were also applied to all deep learning models including the baselines. The deep learning toolkit used in this work is Keras. The loss function to minimize was mean absolute error, and we used Adam \cite{kingma2014adam} as the optimizer. The initial learning rate of optimizer was set to 0.001 and decreased by 0.0005 after training 50 epochs. All configurations were trained for maximum 100 epochs with a batch size of 4 spoken utterances. The model which had the smallest loss on validation set was selected. Silent segments at both ends of utterances were not trained and evaluated. We fixed a maximum length of 710 frames on VTR dataset. Short utterances were padded zeros if they were shorter than the fixed maximal length. During training and testing, we used the Masking layer of Keras to locate the zero time step to be skipped.

\subsection{Baselines}
The LSTM tracking model was trained using the same model configuration from \cite{dissen2019formant} except for the previously mentioned optimizer and loss function. On this basis, we trained the Bi-LSTM tracking model by replacing the LSTM layers with Bi-directional LSTM layers. We further trained the TCN based tracking model using the same model parameter settings for our proposed model. In addition to the three neural network formant tracking models, we also extracted formants using two widely used speech analysis tools: WaveSurfer and Praat.

\subsection{Metrics}
Two quality measures were calculated to quantify the distance of formant tracker output to the annotation reference:

\begin{itemize}
\item MAE: mean absolute error between reference and formant tracker output calculated over speech frames.
\item MAPE: mean absolute percent error between reference and formant tracker
output calculated over speech frames.
\end{itemize}

The smaller the value is, the more the formant tracker output matches the reference.

\section{Results and Discussion}
\begin{figure}[htb]
  \hspace{5pt}
  \includegraphics[width=\linewidth]{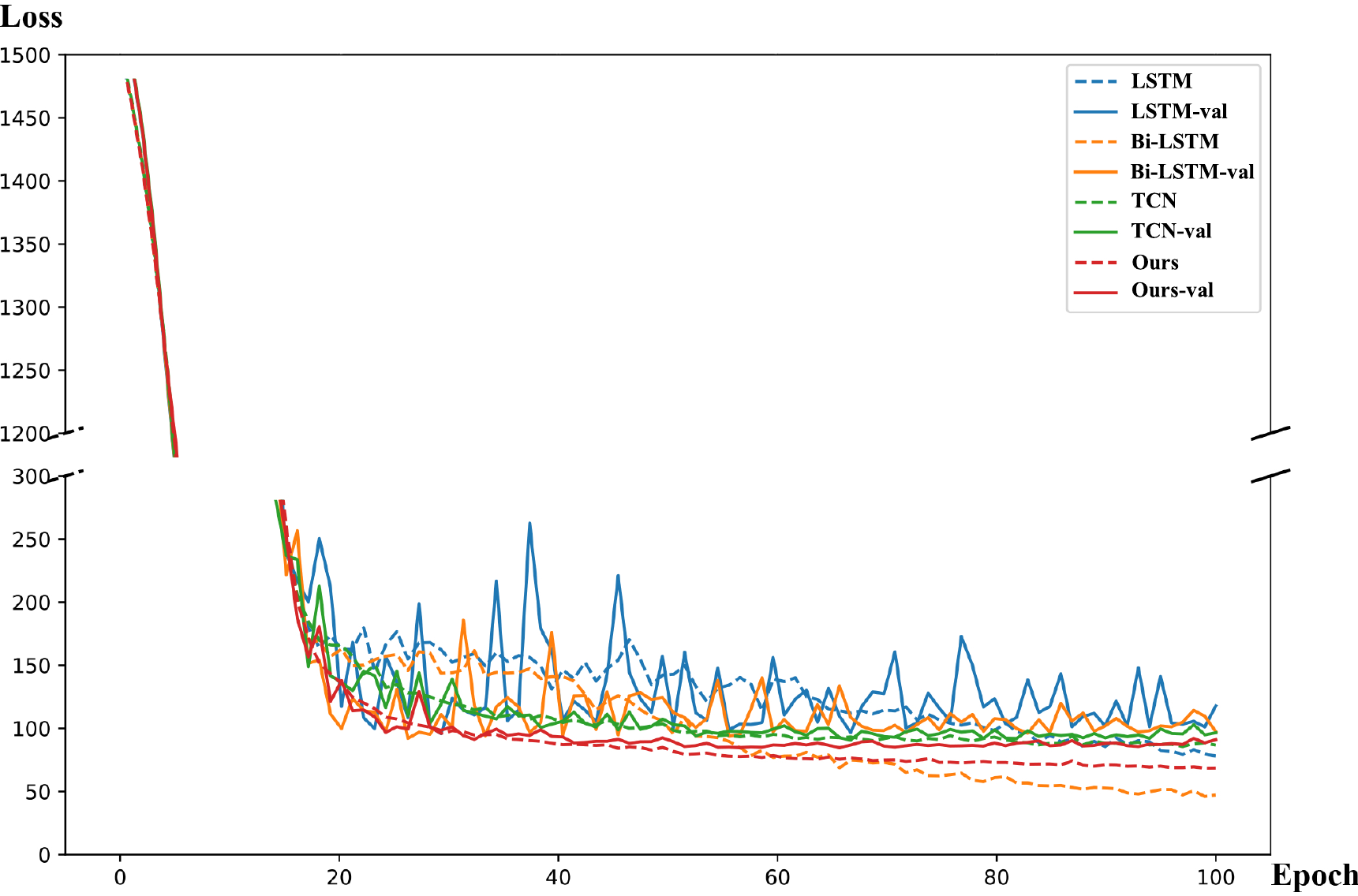}
  \caption{Training and validation loss for different neural network models.}
  \label{fig:compareLoss}
\end{figure}

Figure \ref{fig:compareLoss} shows the training (dashed lines) and validation (solid lines) loss for different neural network models. Although they follow the same trend in the beginning stage, the LSTM based models (curves with obvious fluctuations as shown in Figure \ref{fig:compareLoss}) appear to be more difficult to converge than other models after the first 20 epochs, even over-fitting happened to the Bi-LSTM model. With a faster convergence speed, ``Ours" achieved even better performance than the TCN model.

\begin{figure*}[htb]
  \centering
  \includegraphics[width=\linewidth,keepaspectratio]{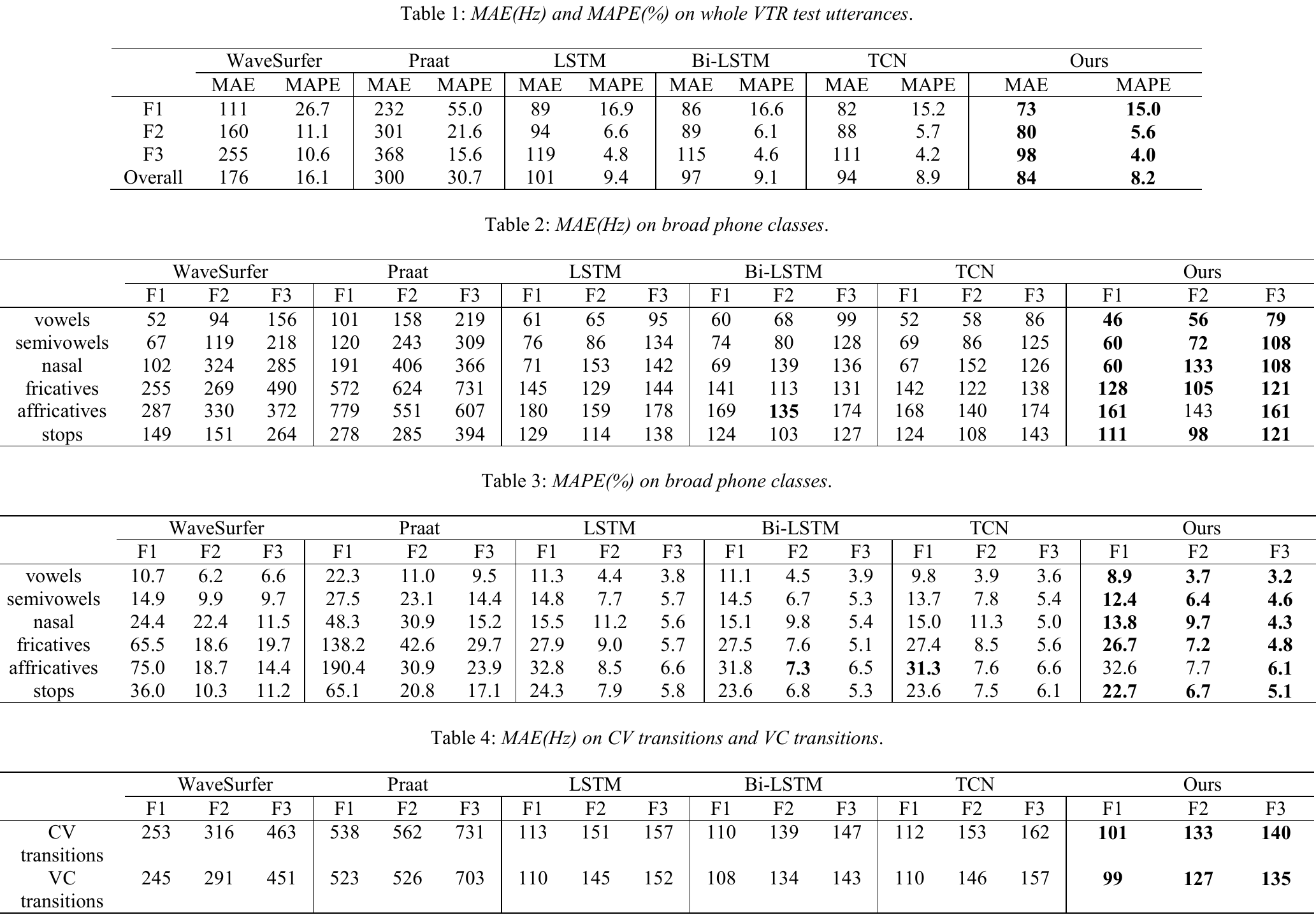}
  \label{fig:compareLoss1}
\end{figure*}

Tables 1-4 present the quantitative error analysis for our model. Different from \cite{dissen2019formant} where they trained models on a subset but testing on the whole dataset, we assured that there was no overlap between the training set and the test set, thus objectively evaluating the generalization performance of supervised model.

Table 1 shows the precision of F1, F2, F3 and overall in MAE and MAPE. (Note that the results were only caculated over speech-labeled frames \cite{mehta2012kalman} for the formant estimation of non-speech is meaningless.) From this table, we can see neural network models/trackers significantly outperformed WaveSurfer and Praat. The results also show the effectiveness of convolutional architectures for formant tracking. It is worth mentioning that our model achieved the smallest error rate even compared with the advanced Bi-LSTM and TCN model. We further categorized the speech sounds to six categories like \cite{dissen2019formant}.

Table 2 and Table 3 respectively presents the accuracy in MAE and MAPE for each broad phone class. The depth formant tracking models outperformed Praat and WaveSurfer almost in every category, except that WaveSurfer had a better estimation of F1 on vowels. The TCN model had a better accuracy than the LSTM and Bi-LSTM model on vowels, semivowels (excepts for F2) and nasal (excepts for F2). The overall best performance on almost every phone was achieved by our proposed model, excepts for F2 of affricatives.

We also examined the errors of the algorithms when limiting the error-counting regions to only the consonant-to-vowel (CV) and vowel-to-consonant (VC) transitions. In this study, the number of frames of the transition regions were not fixed like \cite{dissen2019formant} as the CV or VC boundary was not known in advance in practical application. Our model also achieved the best precision among all methods.

\section{Conclusions}
In this paper, we proposed a novel temporal convolutional network upon the conventional TCN model for formant tracking. The ``causal'' mode of dilated convolution was turned off to capture the impact of speech context. Each layer reused the output from \textit{all} previous layers through the dense connection. With the gating mechanism, the model selectively forgets unimportant information. The approach was validated on an open access dataset. The experimental results showed that our model achieved the best performance on almost all broad phone classes and transitions, compared to LSTM based models and TCN model. In the future, we will consider estimating the formants of vowels segments and investigating whether pre-training is helpful for this task.

\section{Acknowledgements}
This work is supported by National Social Science Foundation of China (18BYY124), Advanced Innovation Center for Language Resource and Intelligence (KYR17005), Discipline Team Support Program of Beijing Language and Culture University (GF201906), Wutong Innovation Platform of Beijing Language and Culture University (19PT04), the Fundamental Research Funds for the Central Universities, and the Research Funds of Beijing Language and Culture University (20YCX158). The corresponding author of the paper is Yanlu Xie.




\end{document}